\documentclass[aps,pra,superscriptaddress,amsmath,amssymb,preprintnumbers,floatfix,showpacs,showkeys,12pt]{revtex4}
\usepackage{amssymb} \usepackage{epsfig} \usepackage{multirow}
\begin{document}
\title{Shor's quantum algorithm using electrons in semiconductor
  nanostructures } \author{Fabrizio~Buscemi }
\email{fabrizio.buscemi@unimore.it} \affiliation{Department of
  Electronics Computer Science and Systems University of Bologna,
  Viale Risorgimento 2, I-40136 Bologna, Italy} \affiliation{ARCES,
  Alma Mater Studiorum, University of Bologna, Via Toffano 2/2, 40125
  Bologna, Italy} \affiliation{Center S3, CNR-Institute of Nanosciences, Via Campi 213A, I-41125 Modena}

\begin{abstract}
  Shor's factoring algorithm illustrates the potential power of
  quantum computation.  Here, we present and numerically investigate a
  proposal for a compiled version of such an algorithm based on a
  quantum-wire network exploiting the potential of fully coherent
  electron transport assisted by the surface acoustic waves.
  Specifically, a non standard approach is used to implement, in a
  simple form, the quantum circuits of the modular exponentiation
  execution for the simplest instance of the Shor's algorithm, that is
  the factorization of $N$=15.  The numerical procedure is based on a
  time-dependent solution of the multi-particle Schr\"odinger
  equation.  The near-ideal algorithm performance and the large
  estimated fidelity indicate the efficiency of the protocol
  implemented, which also results to be almost insensitive to small
  destabilizing effects during quantum computation.

\end{abstract}

\maketitle
 

\section{Introduction}

Quantum computers can efficiently solve some problems that are
unaffordable on classical computers, and processing the information
encoded in quantum systems results to be extremely powerful for
particular tasks.  Specifically, quantum mechanics effects like
entanglement and wavefunction superposition turn out to be fundamental
building blocks, and allow for the quantum computational speedup over
classical computation.  The Shor's
algorithm~\cite{Shor1,Shor2,Beckman,Ekert} has undoubtedly been widely
investigated among those illustrating the power of quantum
computation. As a matter of fact, it plays a key role in cryptographic
protocols, because it allows one to factorize a composite number with
a computational time that is a polynomial function instead of an
exponential function of the number itself.

The practical implementation of the Shor's algorithm represents a
challenge for quantum information science. Two possible physical
architectures have been proposed: nuclear magnetic resonance
(NMR)~\cite{Vandersypen1,Vandersypen2} and
photonic~\cite{Lanyon,Lu,Politi} systems, even though some open
questions exist in both cases.  While in NMR it is difficult to
prepare the qubits in pure states and control their coherent
evolution, thus leading to a controversial quantum nature of the
experiments, photonic systems cannot be scaled to a larger number of
qubits due to their size and stability limitations.  Nevertheless, a
recent experimental demonstration of Shor's algorithm was obtained by
means of optical waveguides integrated on silica-on-silicon chips.
Even if the efficiency of the single photon source and detectors still
does not appear to be very good~\cite{Politi}, the suggested
architecture is promising for the implementation of large-scale
quantum circuits on many qubits.

No evidence of compiled version of quantum factoring algorithm using
electron qubits has been achieved so far. The approach of using charge
carriers in solid-state systems is very appealing because it does not
allow only to overcome the scalability problem, but it also provides a
valid guideline for the design of devices easily integrable in the
traditional electronic circuitry. Specifically, the possibility to
implement Shor's quantum factoring algorithm on an electronic chip
would certainly represent an essential test to verify the potential of
quantum cryptography on everyday life.

In this paper, we propose and numerically simulate a compiled version
of the Shor's algorithm. Electronic quantum logic gates in
one-dimensional (1D) semiconductor channels are used to realize the
necessary processes and to produce multi-particle entanglement and
multipath interference. In particular, we have considered the fully
coherent Surface-Acoustic-Wave (SAW) assisted electron transport in
couples of GaAs quantum wires~\cite{Barnes2}, with the qubit defined
by the localization of a single carrier in one of the coupled
channels~\cite{Rodriquez}.

Quantum-wire systems have been shown to be suitable to produce
bipartite entangled states~\cite{Bertoni2} and to perform quantum
teleportation~\cite{Buscemi2}. Here, the numerical implementation of
the quantum factoring algorithm results to be much demanding in
comparison with the previous works~\cite{Bertoni2,Buscemi2}, due to
the higher number of the simulated quantum logic operations over many
qubits. Specifically, we design the quantum circuits of the modular
exponentiation execution for the easiest meaningful instance of Shor's
algorithm, that is the factorization of $N$=15 for two different
co-primes $C$=11 and $C$=2 (defined in Sec.~\ref{phy}),
corresponding to the period $r$=2 and
$r$=4, respectively. The circuit performing the modular exponentiation
function is brought to a form different from the one given in
literature~\cite{Shor1,Ekert}. This procedure allows one to move on
toward simpler networks of electron quantum gates, and aims at future
research leading to a scalable full-realization of the Shor's
algorithm in quantum-wire devices. In our implementation the inverse
quantum Fourier transformation (QFT) is not present since it is not
necessary for any order-2$^l$ circuit (with $l \in \mathbb{N}$), as
shown in the literature~\cite{Lanyon}. For sake of completeness, a
description of the circuit realizing the inverse QFT in a quantum-wire
network has been given elsewhere~\cite{Reggiani}.

This paper is organized as follows. In Sec.~\ref{phy} we illustrate
the theoretical features of Shor's algorithm, while the description of
the physical implementation in a quantum-wire device and the
discussion of the numerical approach adopted are given in
Sec.~\ref{Phyimp}.  In Sec.~\ref{RESULTS} we show the results obtained
from the numerical simulations of two quantum circuits for the
factorization of $N$=15 corresponding two different parameters choice.
Comments on the results and final remarks are drawn in
Sec.~\ref{Conclu}.

\section{Shor's algorithm}\label{phy}
The strategy to find a nontrivial prime factor of the positive integer
$N$ is described in the following. A random co-prime $C$ is chosen,
i.e. $N$ and $C$ have not common factors. Euler's theorem states that
exists an integer $r$ such that $C^r=1\!\!\!\mod N$ (that is, $C^r-1$
is an integer number multiple of $N$), with $1\le r <N$. The number
$r$ is called \emph{order} of $C \mod N $.  Provided that the latter
is even, then it follows that $C^r-1=(C^{r/2}-1)(C^{r/2}+1)=0\mod N$
and this implies that $N$ is a divisor of the product
$(C^{r/2}-1)(C^{r/2}+1)$.  Assuming that $(C^{r/2}\neq -1 \mod N)$, it
follows that $N$ must have a common factor with both $(C^{r/2}\pm 1)$.
Therefore, this implies that the factors of $N$ are given by the
greatest common divisor of $N$ and $(C^{r/2}\pm 1)$, which can be
efficiently computed by means of Euclid's classical algorithm. It is
worth noting that in order to guarantee the algorithm validity the two
conditions stating that $r$ is even and $(C^{r/2}\neq -1 \!\!\!\mod
N)$ must be satisfied.  These conditions are met with high probability
for $N$ odd, except in the case where $N$ is a prime power
($N=p^{\alpha}$ with $ p $ prime).  Thus, the smallest composite
integer $N$ that can be successfully factored by Shor's method is
$N=15$. If $N$ is an even number or a prime power, other classical
methods should fruitfully be applied for the factorization instead of
the Shor's method.
 
The Shor's algorithm needs quantum computation only in the evaluation
of the order $r$. The quantum order-finding routine commonly uses two
registers of qubits~\cite{Shor1,Beckman}: the argument register with
$n=2\ln_2 N$ qubits, and the function-register with $m=\ln_2 N$
qubits. Its implementation can be separated into three distinct steps.
The first one is the register initialization corresponding to
\begin{eqnarray}
  |0\rangle^{\otimes n}|0\rangle^{\otimes m}  \longmapsto & & \frac{1}{\sqrt{2^n}}
  \left(|0\rangle + |1\rangle \right)^{\otimes n}|0\rangle^{\otimes m-1}|1\rangle=  \nonumber \\ & &\frac{1}{\sqrt{2^n}}\sum_{x=0}^{2^n-1} |x\rangle |0\rangle^{\otimes m-1}|1\rangle,
\end{eqnarray}
where the argument register is prepared by means of the Hadamard
transformations in an equal superposition of all $n$-qubit
computational basis $|0(1)_{x_1}\rangle |0(1)_{x_2}\rangle \cdots
|0(1)_{x_i}\rangle \cdots |0(1)_{x_n}\rangle$. In the second step,
also known as modular exponentiation, the function $C^x \!\!\!\!\mod
N$ is implemented on the function register, while the argument
register remains in $x$. The global state is thus given by
\begin{equation} \label{modexp} \frac{1}{\sqrt{2^n}}\sum_{x=0}^{2^n-1}
  |x\rangle |C^x\!\!\!\!\!\!\mod N\rangle.
\end{equation}
The state of Eq.~(\ref{modexp}) is highly entangled and exhibits the
so-called ``massive parallelism'', i.e. the execution entangles in
parallel all the $2^n$ input values with the corresponding values of
$C^x\!\!\!\!\mod N$, although the algorithm has run only
once~\cite{Beckman,Deutsch}.  Finally, the inverse QFT is applied on
the argument register yielding the state
\begin{equation}
  \frac{1}{2^n}\sum_{z=0}^{2^n-1}\sum_{x=0}^{2^n-1} e^{2\pi xz/2^n} |z\rangle |C^x\!\!\!\!\!\!\mod N\rangle,
\end{equation}                                                            
where, due to the interference, only the terms $|z \rangle$ with
\begin{equation} \label{order} z=a2^n/r
\end{equation}
have a significative amplitude. In the above, $a$ is a random integer
ranging from 0 to $r$-1. Thus, if one performs measurements on the
outcome of the argument register, he would get $a2^n/r$ for some $a$,
and the order $r$ can be deduced after the classical procedure with
probability greater than 1/2 (see Ref.~\onlinecite{Beckman}).

The modular exponentiation, that is the evaluation of $C^x
\!\!\!\!\mod N$ for $2^n$ values of $x$ in parallel, is the most
demanding part of the algorithm. This can be performed by using the
identity: $x=x_{n-1}2^{(n-1)}+\cdots+x_12^1+x_02^0$, where $x_k$ are
the binary digits of $x$. From this, it follows that
\begin{eqnarray} \label{modexpo}
  C^x\!\!\!\!\!\!\mod N=C^{2^{(n-1)}x_{n-1}}\cdots C^{2x_1}C^{x_0}\!\!\!\!\!\!\mod N =\nonumber \\
  C^{2^{(n-1)}x_{n-1}}\cdots \![ C^{2x_1}[C^{x_0}\!\!\!\!\!\!\mod N]
  \!\!\!\!\!\!\mod N]\cdots \!\!\!\!\!\mod N].
\end{eqnarray} 
This means that we first multiply 1 by $C \!\!\!\!\mod N$, if and only
if $x_0=1$; then we multiply the result by $C^2 \!\!\!\!\mod N$ if and
only if $x_1 = 1$ and so forth, until we finally multiply by
$C^{2^{(n-1)}} \!\!\!\!\mod N$ if and only if $x_{n-1} = 1$.
Therefore, the modular exponentiation consists of $n$ serial
multiplications modulo $N$, each of them controlled by the qubit
$x_k$.  The factors $C,C^2,\cdots,C^{2^{(n-1)}}\!\!\!\!\!\mod N$ can
be found efficiently on a classical computer by \emph{repeated
  squaring}.

\subsection{N=15}
As noted in the previous section, the Shor's factorization algorithm
fails if $N$ is even or a prime power, and the smallest composite
integer $N$ that can be successfully factored by means of Shor's
method is $N$=$15$. Even if $N$ is small, this compiled version of the
Shor's algorithm displays a great potential for a future realization
of large-scale quantum algorithm.

Being $N$=$15$, the minimum size of the function and argument
registers must be $m$=$\ln_2{[15]}$=4 and $n$=$2m$=8, respectively.
The algorithm would then require at least 12 qubits.  However, the
following comments allow us to reduce the number of qubits necessary
for the purpose of a proof-of-principle demonstration.  A co-prime $C$
with 15 is one element of the set 2, 4, 7, 8, 11, 13, and 14. As shown
in Table~\ref{tab1}, it comes from repeated squaring that
$C^4\!\!\!\!\mod 15=1$ for all valid $C$. In turn, this implies that
only two bits $x_0$ and $x_1$ are needed for the controlled
multiplications. As a consequence, the multiplications by
$C^4,C^8,\cdots$ are trivial, and all the multiplications, except the
ones by $C$ and $C^2$, can be left out. For $C=4, 11, 14$,
$C^2\!\!\!\!\mod 15=1$ and only the first bit $x_0$ is relevant. These
considerations account for a reduction of the size of the argument
register, which can finally be constituted by no more than two qubits
($n$=2).  Adding this latter result to the four qubits of the function
register, only six qubits are needed instead of twelve, as previously
found.


\begin{table}
  \begin{tabular}{  c c|c c c c c c c }
    & &\phantom{11} &\phantom{11} &\phantom{11}&$C$ &\phantom{11} &\phantom{11} &\phantom{11} \\
    &  &   2& 4 &7  &8  &11&13&14\\    \hline
    & 0& 1& 1&1&1&1&1&1 \\
    \multirow{2}{*}{$x$}&  1 & 2&4&7&8&11&13&14 \\ 
    & 2 & 4&1&4&4&1&4&1 \\
    & 4 & 1&1&1&1&1&1&1 \\ 
  \end{tabular} 
 \caption{ The table displays $C^{x}\!\!\!\!\!\mod 15$ for all $C<15$ co-prime
with 15 and for values of $x$ which are power of two.  Note that $C^{4}\!\!\!\!\!\mod 15=1$ for all valid $C$.   
\label{tab1} }
\end{table}

Shor's factorization algorithm for the number 15 turns out to be
particularly simple because it does not require the implementation of
the inverse QFT in the quantum circuit. As shown in the
literature~\cite{Lanyon}, the latter is not necessary for any circuit
of order $2^l$ and It can be replaced by a classical processing which
also inverts the order of the computed quantum bits of the argument
register.

In this work we implement the Shor's quantum factoring algorithm and
check it against two co-primes: $C$=$11$ and $C$=$2$, that are
representative parameters for the system at hand.

 \subsubsection{$C$=11}\label{c11}
 This parameter choice represent an ``easy case'' since the modular
 exponentiation can be simplified to the multiplication of the initial
 function register state, $|y\rangle$=$| 0_{y_3} 0_{y_2} 0_{y_1}
 1_{y_0} \rangle$=1, by $C$=11 controlled only by
 $x_0$~\cite{Vandersypen2}.
 
\begin{figure}[htbp]
   \centering
   \begin{minipage}{.45\textwidth}
     \includegraphics*[width=\textwidth]{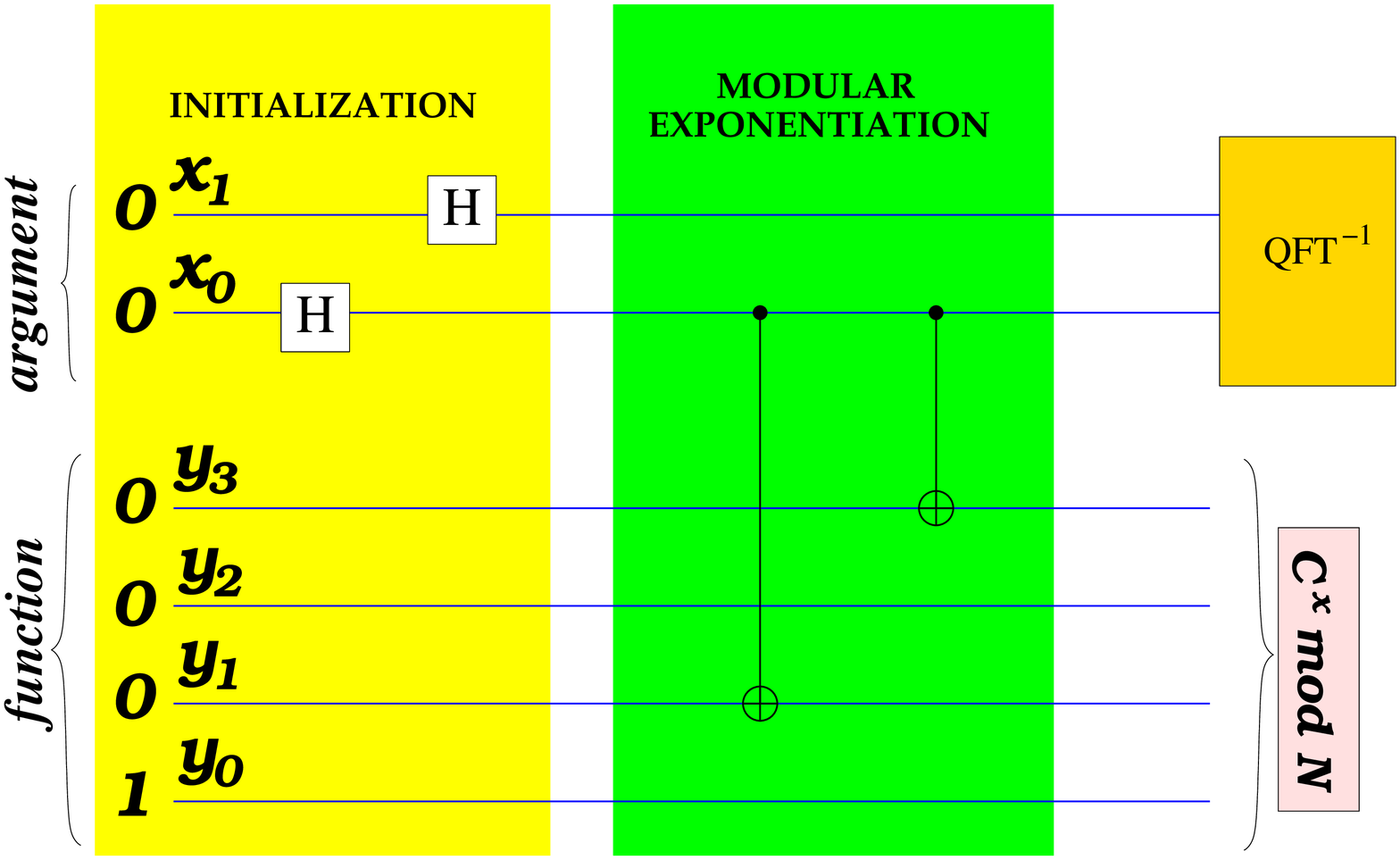}
   \end{minipage}
   \begin{minipage}{.45\textwidth}
     \includegraphics*[width=\textwidth]{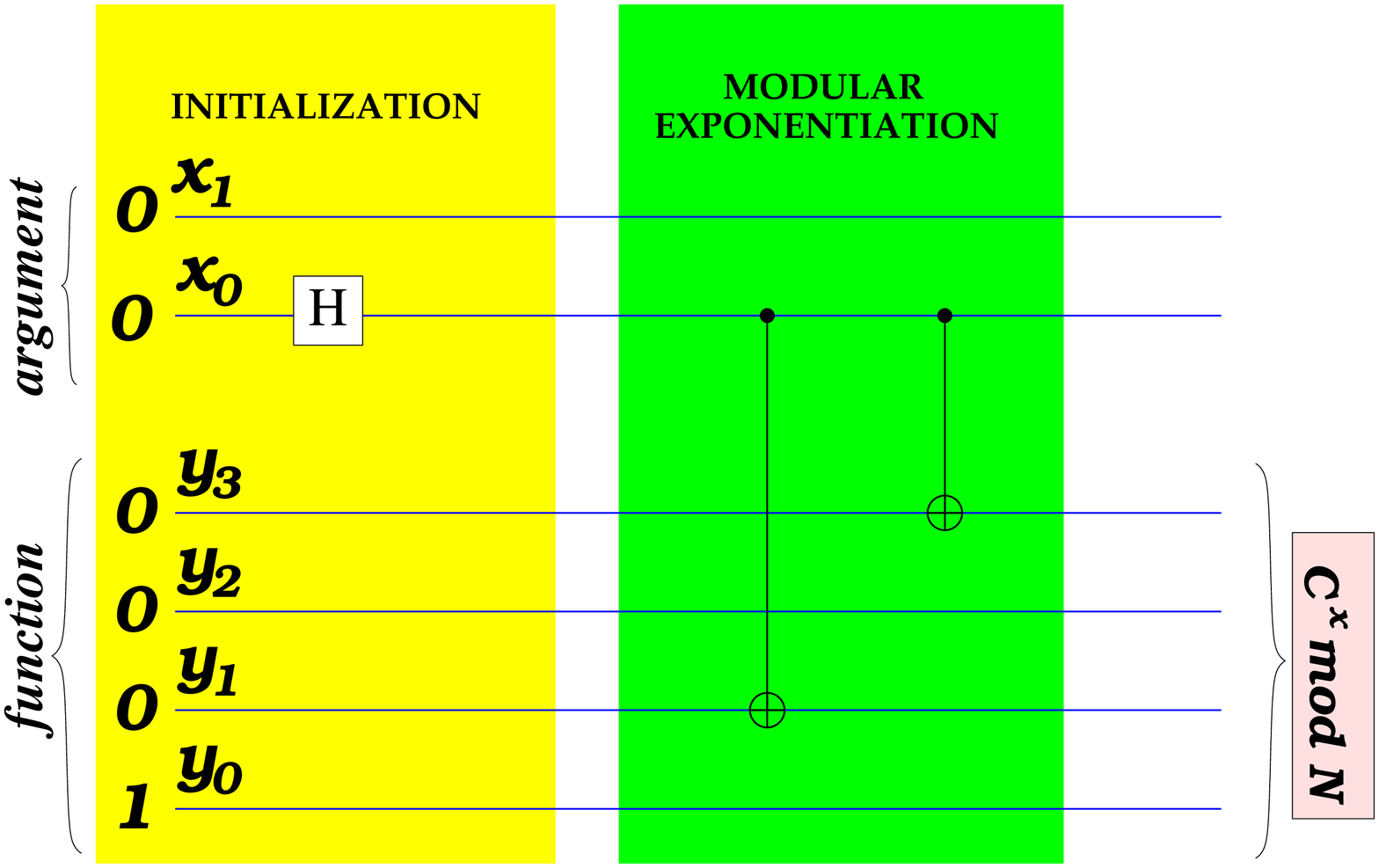}
   \end{minipage}
   \caption{\label{fig4} (Color online).Left panel: Outline of the quantum circuit
     for quantum factorization of 15 for $C$=$11$, using the inverse
     QFT.~\cite{Lu} Right panel: Outline of the quantum circuit for
     quantum factorization of 15 for $C$=$11$, not using the inverse
     QFT.  }
 \end{figure}

 In the left panel of the Fig.~\ref{fig4} a compiled version of the
 quantum circuit for $C$=$11$, using the inverse QFT, is displayed. At
 first, both registers are initialized: each qubit of the argument is
 prepared by Hadamard gates in a superposition of 0 and 1, and the
 function register state is set to $|y\rangle$=$1$, so that the global
 state $|\Phi_{C=11}\rangle$ of the system is
 \begin{eqnarray}
   & & |\Phi_{C=11}\rangle=  \frac{1}{2} \left(|  0_{x_1} 0_{x_0} \rangle +|  0_{x_1} 1_{x_0}  \rangle +| 1 _{x_1} 0_{x_0}  \rangle +|  1_{x_1} 1_{x_0}  \rangle \right)  \nonumber \\
   & &\times| 0_{y_3}  0_{y_2}  0_{y_1}  1_{y_0} \rangle .
 \end{eqnarray}

 Then, the modular exponentiation is performed: the controlled
 multiplication of 1 by 11 is equivalent to the controlled addition of
 10 to 1. The latter is implemented in the quantum circuit by two
 controlled-NOT (CNOT) gates: one between $x_0$ and $y_1$ and one
 between $x_0$ and $y_3$. It is worth noting that the qubits $y_0$ and
 $y_2$ evolve trivially during computation.  Thus the state of the
 system takes the form
 \begin{eqnarray}
   \lefteqn{   |\Phi_{C=11}\rangle=\sum^3_{x=0} |x\rangle |11^x \!\!\!\!\mod15\rangle= {}}\nonumber \\ & &\frac{1}{2} \times\bigg(|  0_{x_1} 0_{x_0} \rangle  | 0_{y_3} 0_{y_2} 0_{y_1} 1_{y_0}\rangle +|  0_{x_1} 1_{x_0} 
   \rangle  | 1_{y_3} 0_{y_2} 1_{y_1} 1_{y_0}\rangle \nonumber \\
   & & +|  1_{x_1} 0_{x_0} \rangle  | 0_{y_3} 0_{y_2} 0_{y_1} 1_{y_0}\rangle +|  1_{x_1} 1_{x_0} \rangle  | 1_{y_3} 0_{y_2} 1_{y_1} 1_{y_0}\rangle \bigg).
 \end{eqnarray}
 This means that a Greenberger-Horne-Zeilinger (GHZ) entangled state
 \begin{equation} \label{GHZ} |\Psi\rangle=\frac{1}{\sqrt{2}} \left( |
     0_{x_0} 0_{y_3} 0_{y_1}\rangle + | 1_{x_0} 1_{y_3} 1_{y_1}\rangle
   \right)
 \end{equation}
 is created between qubit $x_0$ of the argument register and qubits
 $y_3$ and $y_1$ of the function registers.

 The final step is represented by the inverse QFT. The right panel of
 the Fig.~\ref{fig4} shows a compiled version of the quantum circuit
 without using the inverse QFT. Please note that in this case also
 qubit $x_1$ is redundant: the corresponding Hadamard gate results to
 be unnecessary and does not need to be implemented. Here, the
 initial state is $ \frac{1}{\sqrt{2}} \left(| 0_{x_1} 0_{x_0} \rangle
   +| 0_{x_1} 1_{x_0} \rangle \right) | 0_{y_3} 0_{y_2} 0_{y_1}
 1_{y_0} \rangle $ and the modular exponentiation yields
 $\frac{1}{\sqrt{2}}\left(| 0_{x_1} 0_{x_0} \rangle | 0_{y_3} 0_{y_2}
   0_{y_1} 1_{y_0}\rangle +| 0_{x_1} 1_{x_0} \rangle | 1_{y_3} 0_{y_2}
   1_{y_1} 1_{y_0}\rangle \right)$.
 
 The outcomes of the measurement on the inverted argument qubits $x_0$
 and $x_1$ give then 00 or 10 with equal probability. Once this result
 is known, one can obtain the \emph{order} $r$ of $C \mod N$ from
 Eq.~(\ref{order}). While the output 00 corresponds to a failure, the
 output 10 allows one to determine the period $r=2^2/2=2$ and
 represents a successful implementation of the Shor's algorithm.

 \subsubsection{$C$=2}\label{c2}

\begin{figure}[htbp]
  \centering
  \begin{minipage}{0.45\textwidth}
    \includegraphics*[width=\textwidth]{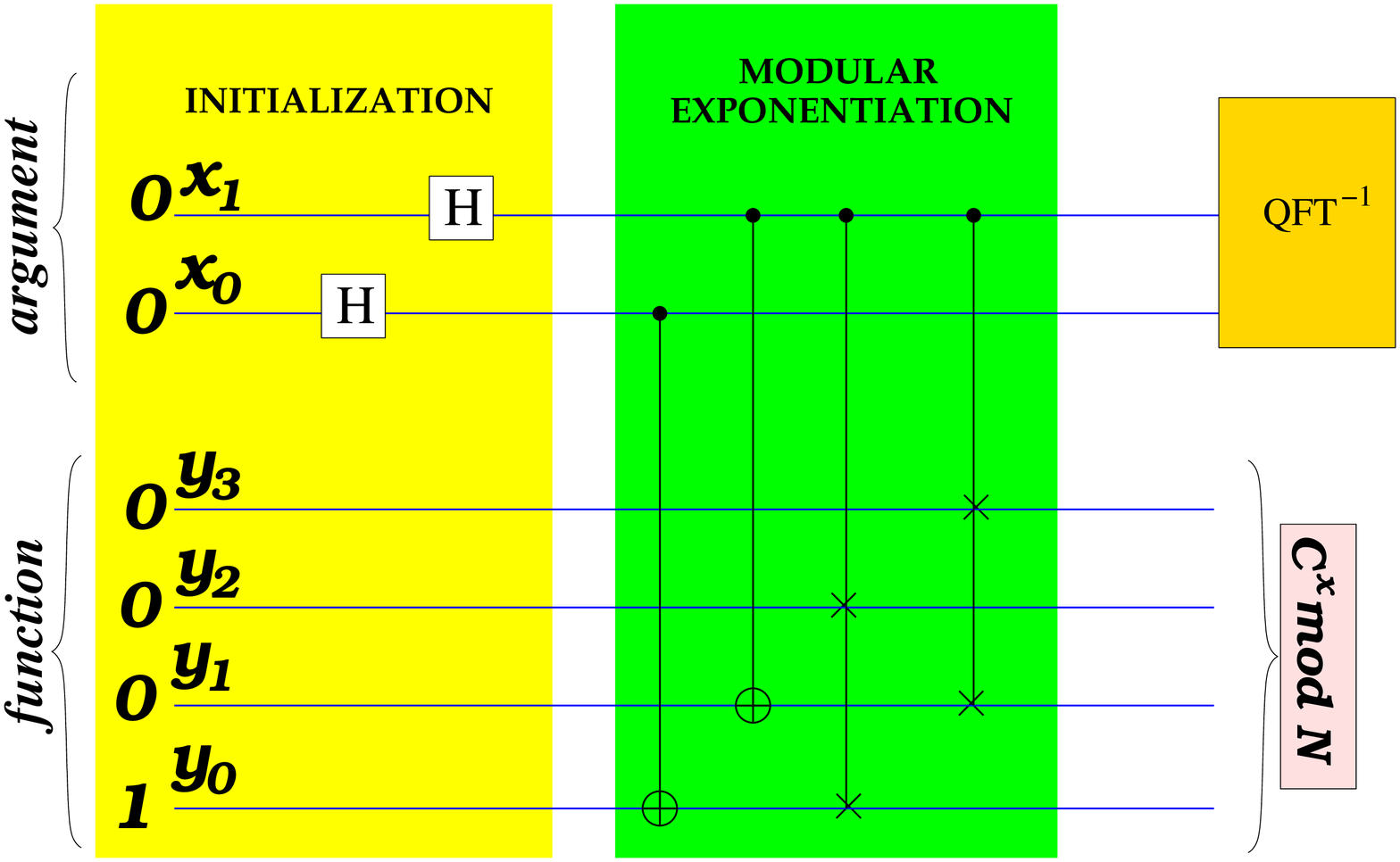}
  \end{minipage}
  \begin{minipage}{0.45\textwidth}
    \includegraphics*[width= \textwidth]{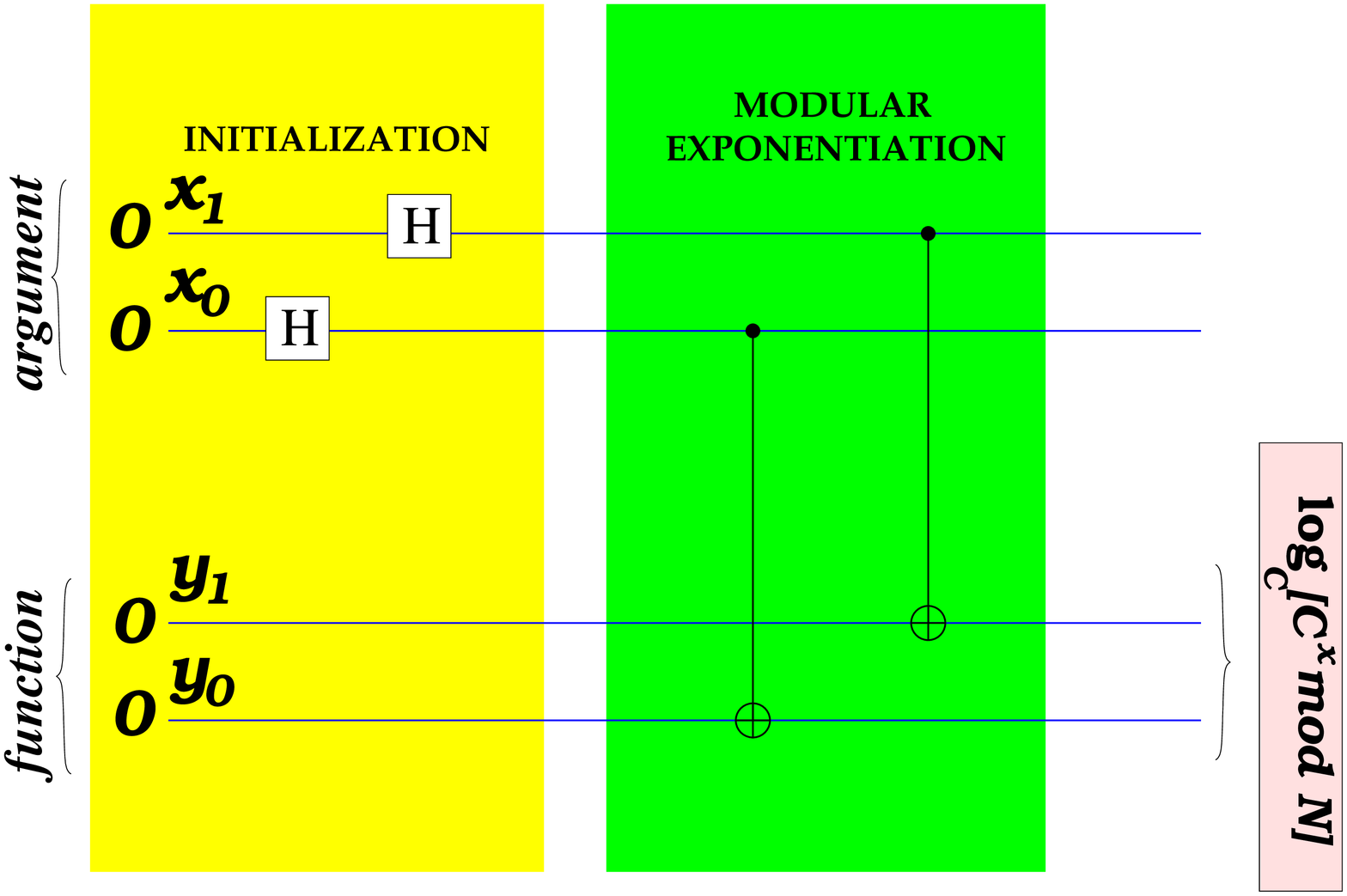}
  \end{minipage}
  \caption{\label{fig2} (Color online).Left panel: Outline of the quantum circuit for
    quantum factorization of 15 for $C$=2, using the inverse QFT and
    evaluating $2^{x}\!\!\!\!\!\mod 15$ in the function register.
    Right panel: Outline of the quantum circuit for quantum
    factorization of 15 for $C$=2, not using the inverse QFT and
    evaluating $\ln_2[2^{x}\!\!\!\!\!\mod 15]$ in the function
    register. Both the circuits correspond to the ones examined in
    Ref.~\onlinecite{Lanyon}.}
\end{figure}

 Since the number of gates needed to perform the modular
 exponentiation is greater than the case of
 $C$=11~\cite{Vandersypen2}, the choice of $C$=$2$ represents a
 ``difficult case''.  In fact, the modular exponentiation is given by
 the multiplication of $|y\rangle$=1 by 2 controlled by $x_0$ and by
 the multiplication of the obtained result by 4 controlled by $x_1$.
 The left panel of Fig.~\ref{fig2} shows the quantum circuit for the
 case at hand. The network for the modular exponentiation is composed
 of two CNOT followed by two controlled SWAP (CSWAP) gates; the first
 two correspond to the addition of 1 to $|y\rangle$=1 controlled by
 $x_0$, while the CSWAP gates multiply the result by 4 controlled by
 $x_1$. The modular exponentiation leads to the state: $ \sum^3_{x=0}
 |x\rangle |2^x \!\!\!\!\mod15\rangle=\frac{1}{2} \times(| 0_{x_1}
 0_{x_0} \rangle | 0_{y_3} 0_{y_2} 0_{y_1} 1_{y_0}\rangle +| 0_{x_1}
 1_{x_0} \rangle | 0_{y_3} 0_{y_2} 1_{y_1} 0_{y_0}\rangle +| 1_{x_1}
 0_{x_0} \rangle | 0_{y_3} 0_{y_2} 1_{y_1} 0_{y_0}\rangle +| 1_{x_1}
 1_{x_0} \rangle | 1_{y_3} 0_{y_2} 0_{y_1} 0_{y_0}\rangle )$.

 Nevertheless, a different compilation of the quantum circuit can be
 realized~\cite{Lanyon}, and it is reported in the right panel of
 Fig.~\ref{fig2}. By means of the latter, it is possible to evaluate
 $\ln_2{[2^x\!\!\!\!\mod15]}$ in the function register in place of
 $2^x\!\!\!\!\mod15$, thus reducing the number of function qubits from
 $\ln_2{[15]}$=4 to $\ln_2{\{\ln_2{[15]}\}}$=2. This compilation
 maintains all the features of the algorithm originally
 proposed~\cite{Beckman}, and still does not make use of the inverse
 QFT, as in the previous case. Following this scheme, the
 initialization of the system leads to the state
 \begin{eqnarray}
   & & |\Phi_{C=2} \rangle =\frac{1}{2} 
   \bigg(|  0_{x_1} 0_{x_0} \rangle +|  0_{x_1} 1_{x_0} \rangle\nonumber \\ & & +| 1 _{x_1} 0_{x_0}  \rangle +|  1_{x_1} 1_{x_0}  \rangle \bigg)      | 0_{y_1} 0_{y_0}\rangle ,
 \end{eqnarray}
 meaning that the argument register is kept in the usual
 equally-weighted coherent superposition of all possible arguments,
 while the initial function register state is $|y \rangle$=0. If we
 apply the procedure described in Eq.~(\ref{modexpo}) to evaluate
 $\ln_2{[2^x\!\!\!\!\mod15]}$, it can be easily shown that the modular
 exponentiation reduces to the sum of $\ln_2{[2^1\!\!\!\!\mod15]}$=1
 to $|y \rangle$=0 controlled by $x_0$, and of
 $\ln_2{[2^2\!\!\!\!\mod15]}$=2 to the obtained result controlled by
 $x_1$.  These operations are implemented in the quantum circuit
 reported in the right panel of Fig.~\ref{fig2} by means of two CNOT
 gates: one between $x_0$ and $y_0$ and another between $x_1$ and
 $y_1$. It is worth noting that in this case the algorithm is very
 simple since it consists of only two networks of gates acting on
 independent qubit pairs.  After modular exponentiation the state of
 the system takes the form
 \begin{eqnarray}\label{Bell}
   |\Phi_{C=2} \rangle =\frac{1}{2}\bigg(|  0_{x_1} 0_{x_0} 0_{y_1} 0_{y_0} \rangle +|  0_{x_1} 1_{x_0} 0_{y_1} 1_{y_0} \rangle \nonumber \\
   +|  1_{x_1} 0_{x_0} 1_{y_1} 0_{y_0} \rangle +|  1_{x_1} 1_{x_0} 1_{y_1} 1_{y_0} \rangle \bigg) \nonumber \\
   =\frac{1}{2}\bigg(|  0_{x_1} 0_{y_1} \rangle +| 1_{x_1} 1_{y_1}
   \rangle \bigg)\bigg(|  0_{x_0} 0_{y_0} \rangle +| 1_{x_0} 1_{y_0} \rangle
   \bigg),
 \end{eqnarray}
 that is the product of two entangled Bell pairs, thus confirming the
 manifestation of entanglement between the two registers of the
 algorithm. The inverse QFT is not necessary and it can be replaced by
 its classical counterpart, which also swaps the output quantum bit of
 the argument register. The two-bit outputs for the case under
 investigations are: 00, 01,10, and 11. The second and the fourth
 outcomes allow the evaluation of the order $r$=4, which efficiently
 yields the factors 3 and 5 via Euclid's classical algorithm; the
 first one corresponds to a failure mode and, lastly, the third one
 leads to trivial factors.

\section{The physical implementation and the numerical
  approach}\label{Phyimp}

Here, we describe the implementation of the Shor's algorithm in
a specific semiconductor nanostructure.  It consists of a number
of couples of GaAs quantum wires where surface acoustic waves (SAWs),
i.e. sinusoidal piezoelectric potential, propagate and trap charged
carriers into their moving minima, letting one particle fill in each
minimum~\cite{Shilton}. The so-called flying qubits are realized by
means of the states $|0\rangle$ and $|1\rangle$, encoded through the
localization of a single electron in one of the two 1D
channels~\cite{Rodriquez}. Here, the SAWs are used to inject and drive
the electron thanks to their efficiency in preventing the natural
spatial spread of the wavefunction~\cite{Barnes1} and in making the
carriers more immune to the decohering effects~\cite{Buscemi}.
Moreover, in this investigation the carrier transport is assumed to be
fully coherent.

As shown in the literature~\cite{Ion,Bertoni}, such a system is able
to provide the universal set of gates useful to realize any quantum
computational network. Specifically, the basic building blocks are
$R_x(\theta)$, $R_{0(1)}(\phi)$, and $T(\gamma)$~\cite{Barenco}. The
former two gates implement one-qubit logical operations, whereas the
latter is a two-qubit gate.

$R_x(\theta)$ acts as an electronic beam
splitter and can be materialized through a coupling window between the
two wires of the qubit~\cite{Bird}. Its matrix representation on the
basis $\{|0\rangle$,$|1\rangle\}$ is given by
\begin{equation}\label{RX}
  R_x(\theta)= \left( \begin{array}{cc}
      \cos{\frac{\theta}{2}} & i\sin{\frac{\theta}{2}} \\
      i\sin{\frac{\theta}{2}} & \cos{\frac{\theta}{2}}
    \end{array}\right) .
\end{equation}

$R_{0(1)}(\phi)$ is an electronic phase shifter obtained by inserting
a potential barrier in the wire 0(1), thus inducing a delay phase
$\phi$ in the propagation of wavefunction. Its action is described in
one-qubit basis by
\begin{equation}
  R_{0}(\phi)= \left( \begin{array}{cc}
      e^{i\phi} & 0 \\
      0 & 1
    \end{array}\right) \qquad \textrm{and}\qquad  R_{1}(\phi)= \left( \begin{array}{cc}
      1& 0 \\
      0 & e^{i\phi}
    \end{array}\right).
\end{equation}

$T(\gamma)$ is a conditional phase gate exploiting the Coulomb
interaction between two electrons. It consists of a region in which
the carriers propagate along two different wires close enough to give
rise to an effective interaction able to delay both particles. The
matrix representation of $T(\gamma)$ in the two-qubit basis $\{|0
0\rangle$,$|0 1\rangle$, $|1 0\rangle$, $| 11\rangle\}$ is:
\begin{equation}
  T(\gamma)= \left( \begin{array}{cccc}
      1 & 0 & 0 & 0 \\
      0 & 1 & 0 & 0  \\
      0 & 0 & e^{i\gamma} & 0 \\
      0 & 0 & 0 & 1
    \end{array}\right) .
\end{equation}
The phases $\theta$, $\phi$, and $\gamma$ of the above quantum gates
depend upon the physical and geometrical parameters of the systems
such as velocity, amplitude and wavelength of the SAW potential,
strength of the electron-electron interaction, coupling window length,
and shape of the potential barrier.  In order to perform any
transformation of the many-qubit state, an appropriate tuning of the
above parameters in a given network of $R_x(\theta)$,
$R_{0(1)}(\phi)$, and $T(\gamma)$ gates is required.

In the compiled versions of the Shor's algorithm illustrated in the
right panels of Figures~\ref{fig4} and \ref{fig2}, corresponding to
$C$=11 and $C$=2, respectively, the two logical operations involved
are only the Hadarmad $H$ and CNOT gates. In terms of $R_x(\theta)$,
$R_{0(1)}(\phi)$, and $T(\gamma)$, these operations can be reworked
as~\cite{Reggiani}:

 \begin{equation}
   H=R_{0}\left(\frac{3\pi}{2}\right)R_{x}\left(\frac{\pi}{2}\right)R_{0}\left(\frac{\pi}{2}\right)R_{1}\left(\pi\right)
   = \frac{1}{\sqrt{2}}\left( \begin{array}{cc}
       1 & 1 \\
       1 & -1 
     \end{array}\right) ,
 \end{equation}
 and
 \begin{equation}
   \mathrm{CNOT}=R^{(2)}_{0}\left(\frac{3\pi}{2}\right)R^{(2)}_{x}\left(\frac{3\pi}{2}\right)
   T^{(1,2)}(\pi)R^{(2)}_{x}\left(\frac{\pi}{2}\right)R^{(2)}_{0}\left(\frac{\pi}{2}\right)R_{1}^{(1)}\left(\pi \right)
   = \frac{1}{\sqrt{2}}\left( \begin{array}{cccc}
       1 & 0 & 0 & 0 \\
       0 & 1 & 0 & 0  \\
       0 & 0 & 0 & 1 \\
       0 & 0 & 1 & 0
     \end{array}\right) .
 \end{equation}

 The numerical implementation of the quantum circuits described in the
 previous section results extremely challenging due to the large
 number of basic building blocks needed to realize the $H$ and
 $\textrm{CNOT}$ gates. Furthermore, the experimental realization of
 such devices would also certainly meet serious obstacles stemming
 from the difficulty of preserving the coherent evolution of the
 qubits during a number of logical operations, from decoherence
 effects due to interactions with the environment, as well as from
 possible structural defects induced by the processes of fabrication
 and tuning of each quantum gate.  The theoretical and experimental 
 implementation of two-qubit quantum circuits by means of a 
 minimum amount of operations has been widely discussed in 
 literature~\cite{Rez, Blaau}.
 Different protocols have been proposed ranging from  the 
 special perfect entanglers~\cite{Rez} to the use of a given tunable 
 entangling interaction~\cite{Blaau}. Here, we propose in the 
 followings a scheme suitable to
 perform the quantum factoring algorithm in devices formed by
 semiconductor quantum wires with a minimal number of the fundamental
 gates $R_x(\theta)$, $R_{0(1)}(\phi)$, and $T(\gamma)$.  The proposed
 implementation satisfies the main requirements of the Shor's
 algorithm as originally formulated~\cite{Shor1, Beckman}. In fact,
 ``massive parallelism'' is maintained, since entanglement is created
 between argument and function registers, and the binary output of the
 argument qubits are unchanged, as it will be shown below. In the left
 and right panel of the Fig.~\ref{fignetwork} we report the
 quantum-wire networks implementing the circuits displayed in the
 right panels of Figures~\ref{fig4} and \ref{fig2}, respectively.
 \begin{figure}
   \begin{center}
     \begin{minipage}[c]{0.45\textwidth}  

       \includegraphics*[width=\textwidth]{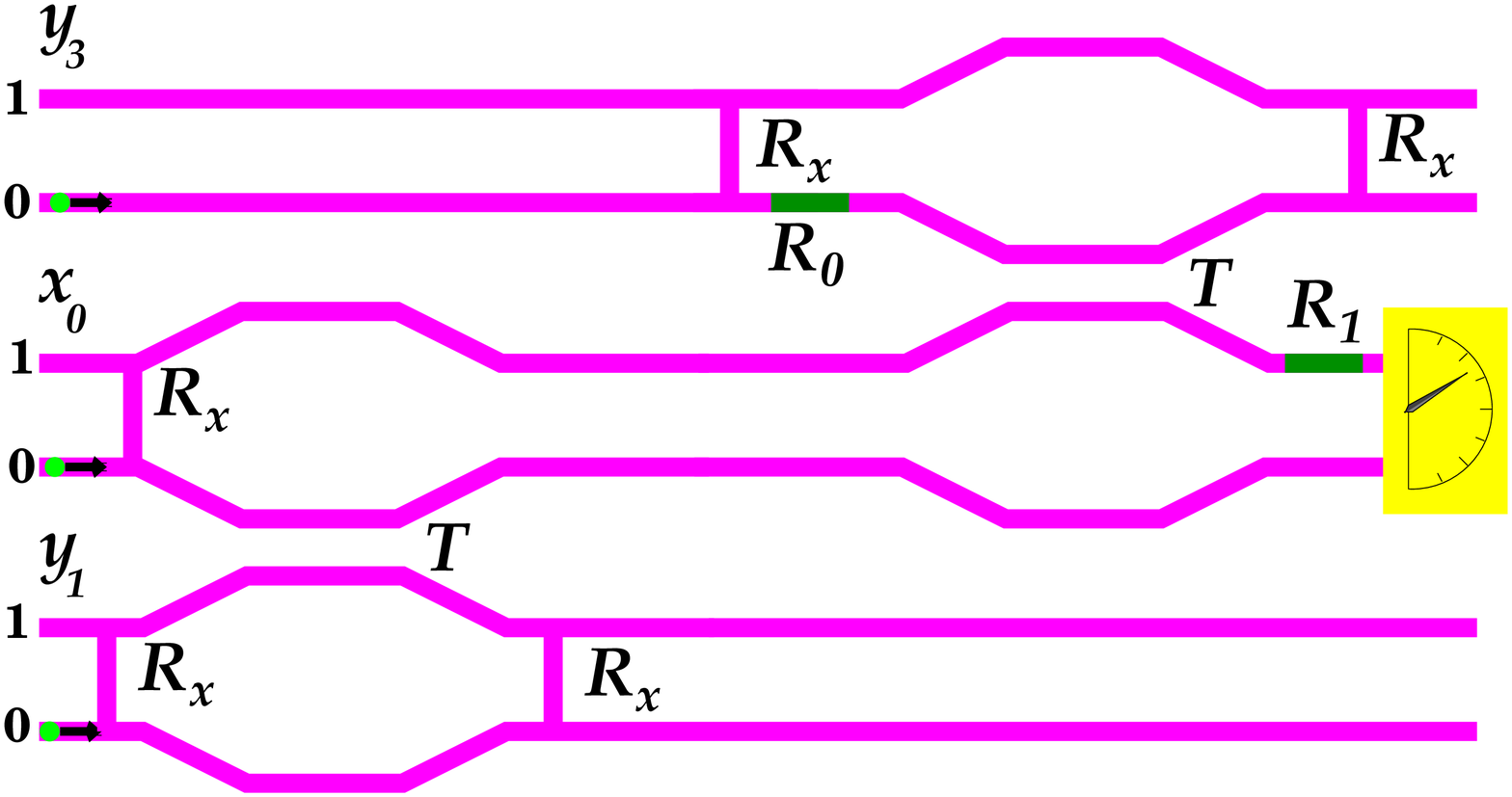}
     \end{minipage}
     \begin{minipage}[c]{0.45\textwidth}
       \includegraphics*[width=\textwidth]{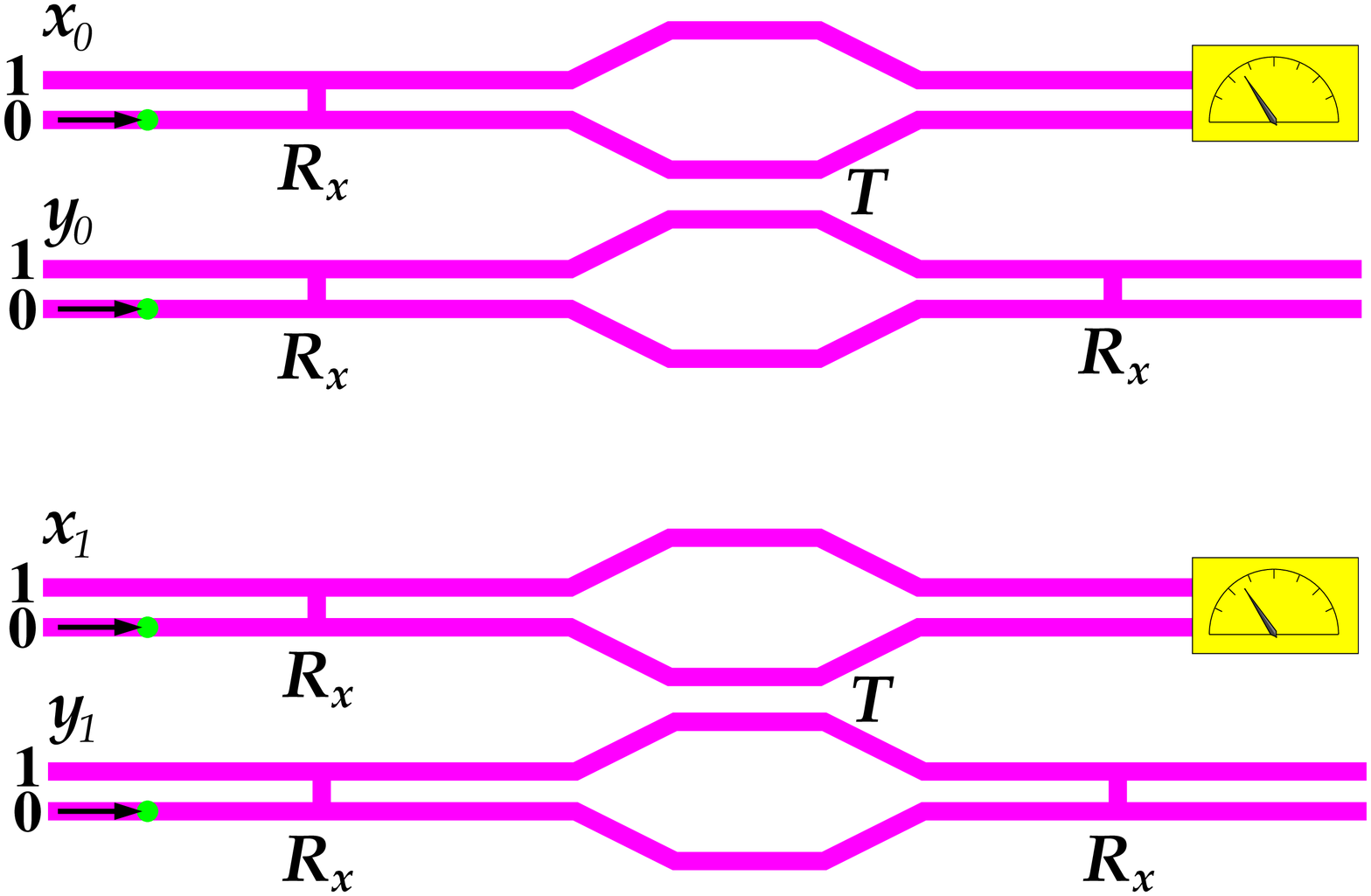}
     \end{minipage}
     \caption{\label{fignetwork}(Color online). Left panel: Sketch of the physical
       system used to factorize $N$=15 with $C$=11, corresponding to
       the quantum circuit displayed in the right panel of
       Fig.~\ref{fig4}.  The beam-splitter $R_x$ of the qubit $x_0$
       mimics the initialization procedure, that is the equal
       splitting of electron wavefunction of the qubit $x_0$ in the
       two wires. The next two sets of gates, first $R_xTR_x$ acting
       on $\{x_0,y_1\}$ and then $R_xR_0TR_1R_x$ on $\{x_0,y_3\}$,
       play the role of two CNOT gates creating the maximum
       entanglement of the qubits $\{x_0,y_1,y_3\}$ in the modular
       exponentiation step.  For sake of clarity the qubits $y_0$ and
       $y_2$ evolving trivially during the computation have not been
       reported. Right panel: Sketch of the physical system used to
       factorize $N$=15 with $C$=2, corresponding to the quantum
       circuit displayed in the right panel of Fig.~\ref{fig2}. In the
       initialization procedure two gates $R_x$ act on the the couple
       of the argument register qubits $\{x_0,x_1\}$ and split each of
       them in an equal superposition of the $|0\rangle$ and
       $|1\rangle$ states. The modular exponentiation consists of two
       network $R_xTR_x$, each operating onto $\{x_0,y_0\}$ and
       $\{x_1,y_1\}$ mimicking the action of two CNOT gates.  Note
       that here, for brevity, the phases of the quantum gates
       involved in the networks and explicitly indicated in the
       Eqs.~(\ref{net1}) and~(\ref{net2}), are omitted.  }
    
   \end{center}
 \end{figure}

 In the first case, corresponding to $C$=11, the network implemented
 reads
 \begin{equation} \label{net1}
   R_{x}^{(x_0)}\left(\frac{\pi}{2}\right)R_{x}^{(y_1)}\left(\frac{\pi}{2}\right)T^{(x_0,y_1)}(\pi)R_{x}^{(y_1)}\left(\frac{\pi}{2}\right)
   R_{x}^{(y_3)}\left(\frac{\pi}{2}\right)R_{0}^{(y_3)}\left(\pi\right)T^{(y_3,x_0)}(\pi)R_{1}^{(x_0)}\left(\pi\right)R_{x}^{(y_3)}\left(\frac{\pi}{2}\right),
 \end{equation}
 where the superscripts of the quantum gates indicate which qubit they
 act on. The three-qubit output state is:
 \begin{equation}\label{GHZ2} \frac{1}{\sqrt{2}} \left(
     -|0_{x_0} 0_{y_3} 0_{y_1}\rangle +i | 1_{x_0} 1_{y_3} 1_{y_1}\rangle
   \right).
 \end{equation}
 For $C$=2 the global logical transformation can be written as:
 \begin{equation}\label{net2}
   R_{x}^{(x_0)}\left(\frac{\pi}{2}\right)R_{x}^{(y_0)}\left(\frac{\pi}{2}\right)R_{x}^{(x_1)}\left(\frac{\pi}{2}\right)R_{x}^{(y_1)}\left(\frac{\pi}{2}\right)T^{(x_0,y_0)}(\pi)T^{(x_1,y_1)}(\pi)R_{x}^{(y_0)}\left(\frac{\pi}{2}\right)R_{x}^{(y_1)}\left(\frac{\pi}{2}\right)
 \end{equation}
 and, after being applied to the input state $| 0_{x_1} 0_{x_0}
 0_{y_1} 0_{y_0}\rangle$, it yields
 \begin{equation}
   \frac{1}{2}\bigg(|  0_{x_1} 0_{y_1} \rangle -| 1_{x_1} 1_{y_1}
   \rangle \bigg)\bigg(|  0_{x_0} 0_{y_0} \rangle -| 1_{x_0} 1_{y_0} \rangle
   \bigg). \end{equation}
 In both  cases, the degree of the entanglement  created between  the qubits of the argument and the function register is equal to the one of the standard formulation of Shor's quantum factoring algorithm~\cite{Shor1, Beckman}. While for $C$=11 the output state is GHZ-like, i.e. the maximum amount of quantum correlations is built up among three qubits, when $C$=2 the four qubit-state is  the product of two entangled Bell pairs. Furthermore, the  reduced density matrices  of the argument qubits correspond to the ones calculated from the quantum states of Eqs.~(\ref{GHZ}) and~(\ref{Bell}). In turn,  this implies  that the binary output of the algorithm, i.e. the outcome measurements on the argument register, be identical.


 The networks of the gates described in Eqs.~(\ref{net1})
 and~(\ref{net2}) have been simulated by solving numerically the
 time-dependent Schr\"odinger equations for three and four electrons
 injected in the GaAs quantum wire devices of Fig.~\ref{fignetwork}.
 While for $C$=2 the four-particle dynamics reduces to the time
 evolution of a pair of separable two-particle systems, when $C$=11, a
 solution of the Schr\"odinger equation for the whole three-carrier
 wavefunction is required~\cite{NOTE3}.  Anyway, in order to minimize
 the computational burden of both of the reported cases, a semi-1D
 approach has been used to investigate the time evolution of the
 system in place of a two-dimensional (2D) computational scheme. This
 approach was already introduced to simulate a teleportation protocol
 in a quantum wire device~\cite{Buscemi2}. According to this
 simplified scheme, the direction $\bf{y}$ of the particles along the
 wires is fully included in the simulation ($\bf{y}$ being discretized
 with a point-grid resolution of $\Delta\bf{y}$=1nm), while the two
 variables describing the longitudinal direction $\bf{x}$ and
 identifying the wire where the carriers are localized can only assume
 one of the two values 0 or 1.

 Though it does not allow to simulate the gate $R_x(\theta)$, which
 would require a full 2D analysis, the numerical procedure here
 adopted makes it possible to move from a time dependent Schr\"odinger
 equation for a multi-variable wavefunction $\Phi(\bf{X},\bf{Y},t)$
 (seven unknowns for the device with $C$=11 and nine when $C$=2) to
 many coupled Schr\"odinger equations of the following kind:
 \begin{equation} \label{Sch} i\hbar \frac{\partial}{\partial
     t}\Phi_{\mathbf{X}}(\mathbf{Y},t)=
   \left(-\frac{\hbar^2}{2m}\nabla^2_{\mathbf{Y}}+
     V_{\mathbf{X}}(\bf{Y},t)\right) \Phi_{\mathbf{X}}(\mathbf{Y},t),
 \end{equation}
 where $\mathbf{X}\equiv
 (\mathbf{x}_{x_{0}},\mathbf{x}_{y_{0}},\mathbf{x}_{x_{1}},\mathbf{x}_{y_{1}},\ldots)$
 and $\mathbf{Y}\equiv
 (\mathbf{y}_{x_{0}},\mathbf{y}_{y_{0}},\mathbf{y}_{x_{1}},\mathbf{y}_{y_{1}},\ldots)$.
 Specifically, when $C$=11 a system of eight coupled equations is
 obtained, while for $C$=2 two independent systems of four equations
 are found since the qubit pairs $\{x_0,y_0\}$ and $\{x_1,y_1\}$ are
 independent. In both the cases a Crank-Nicholson finite difference
 scheme~\cite{NumRec} has been applied to get a numerical solution.
 The potential term $V_{\mathbf{X}}(\bf{Y},t)$ appearing in
 Eq.~(\ref{Sch}) sums up the SAW time dependent potential, the Coulomb
 interaction between electrons, and the static potential profile. The
 simulations here presented make use of a sinusoidal potential
 mimicking a SAW of amplitude and wavelength equal to 20 meV and 200
 nm, respectively, and propagating with the sound velocity
 $v_s=3.3\times 10^3$ m s$^{-1}$.  Screening effects have been
 included in the Coulomb potential between the carriers by inserting
 an exponential damping term~\cite{Ferry} with a Debye wave vector of
 0.2 nm$^{-1}$.

 In order to numerically implement the networks described in
 Eqs.~(\ref{net1}) and~(\ref{net2}), one must firstly to find the
 suitable geometrical parameters of the device for the gates
 $R_{0(1)}(\phi)$ and $T(\gamma)$ giving the required value $\pi$ of
 the phases $\phi$ and $\gamma$.  To this aim, we have performed a
 number of simulations testing different geometries for the phase
 shift and the conditional phase gate. As to the phase $\phi$ is
 concerned, it depends on the height and the length of the potential
 barrier. The values of these parameters obtained from the
 optimization procedure are 2.82 meV and 8 nm, respectively, and
 correspond to a delay phase $\phi$ of 0.92$\pi$, that is good enough
 for our purposes, as it will be shown in the next section. It is
 worth noting that the barrier height is significantly smaller than
 the amplitude of the SAW potential, this making the spatial spreading
 of the electron wavepacket negligible and letting it be entirely
 transmitted through the barrier. The main geometrical parameters
 affecting the phase $\gamma$ of the conditional phase gate are the
 length of the coupling region and the distance between the coupled
 wires. Their optimal values used in the numerical implementation of
 the algorithm are 150 nm and 5 nm, respectively. They lead to a
 $\gamma$ value about 0.88$\pi$, which allows us to simulate
 satisfactorily the two-qubit logical operation of a CNOT-like gate.
 By applying the described geometry for $T(\gamma)$, both the
 tunnelling effects between the two wires and the reflection phenomena
 in the coupling region are negligible.

 From a computational point of view, the numerical simulation of the
 $T(\gamma)$ gate is more challenging than that of the phase shift
 $R_{0(1)}(\phi)$. While the latter involves a one-particle potential,
 the former exploits a two particle-interaction that builds up an
 amount of quantum correlations between the wire degrees of freedom of
 the particles, as expected. However, it creates also an undesired
 entanglement between the variables defining the position of the
 carriers along the wires. As a consequence, the evaluation of the
 effects of the controlled phase gate on the multi-particle
 wavefunction is a demanding task because it implies that a number of
 two particle simulations must be combined together to obtain the
 time-evolved state of the overall system.

 The $R_x(\theta)$ gate has not directly been simulated; nevertheless,
 its action has been taken into account by means of the transformation
 matrix of Eq.~(\ref{RX}), validated by the results of appropriate 2D
 simulations~\cite{Bertoni}.
  
 \section{Results }\label{RESULTS}
 According to the sectioning of the previous paragraphs, we firstly
 present the results obtained for $C$=11, then those for $C$=2.
 \subsection{C=11}

\begin{figure}
   \begin{center}
     \begin{minipage}[c]{0.3\textwidth}  

       \includegraphics*[width= \textwidth]{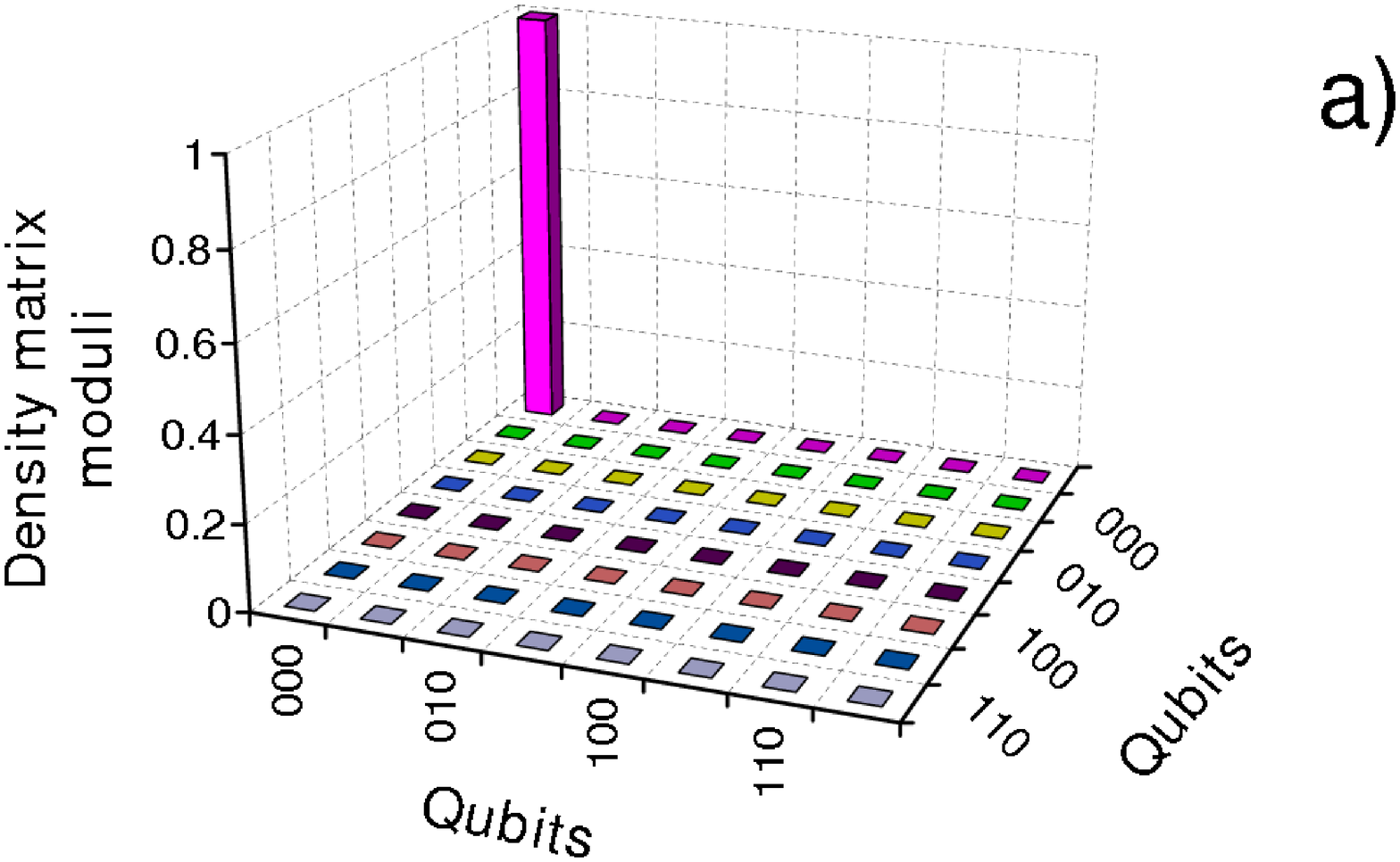}
     \end{minipage}
     \begin{minipage}[c]{0.3\textwidth}
       \includegraphics*[width=\textwidth]{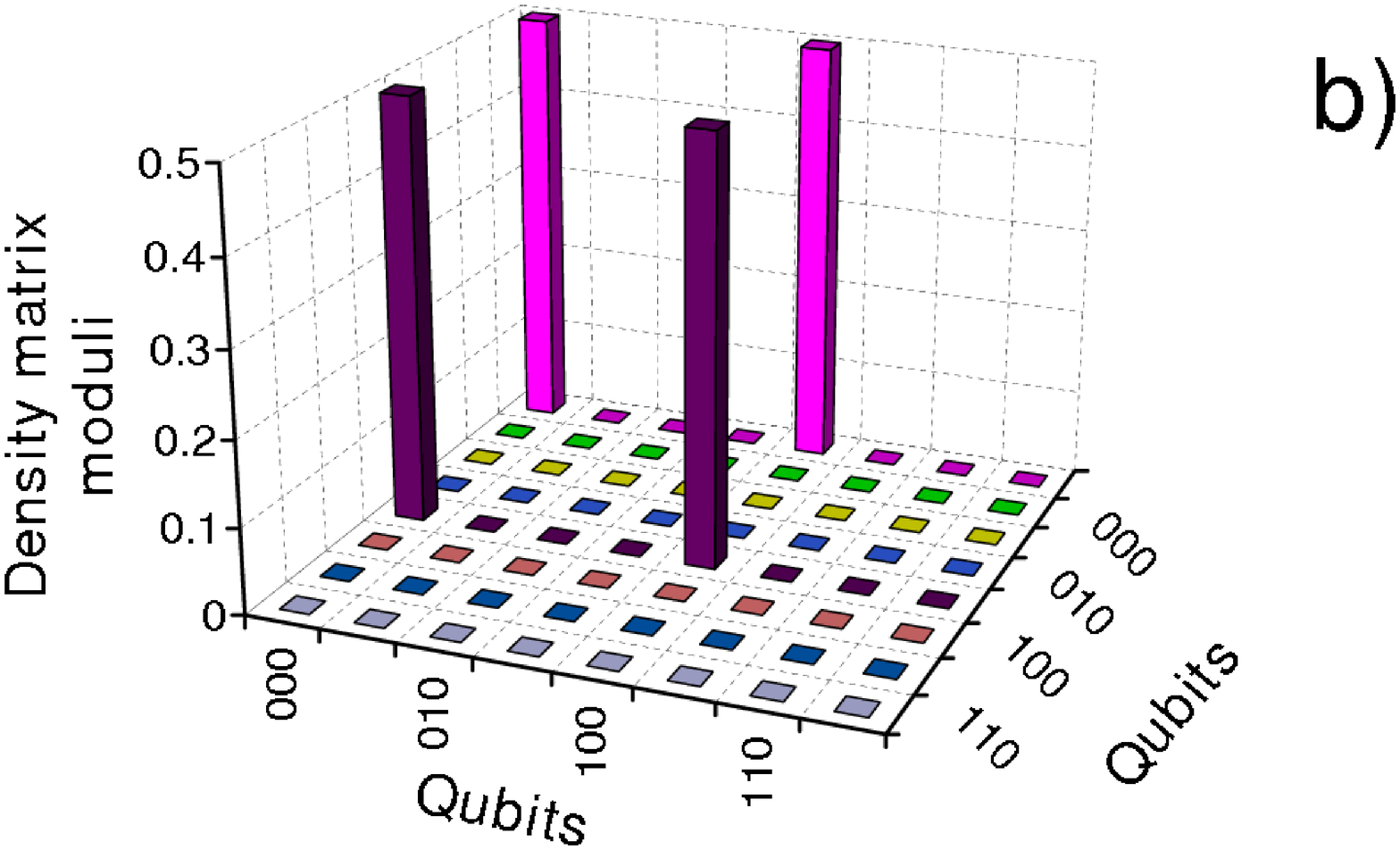}
     \end{minipage}
     \begin{minipage}[c]{0.3\textwidth}
       \includegraphics*[width= \textwidth]{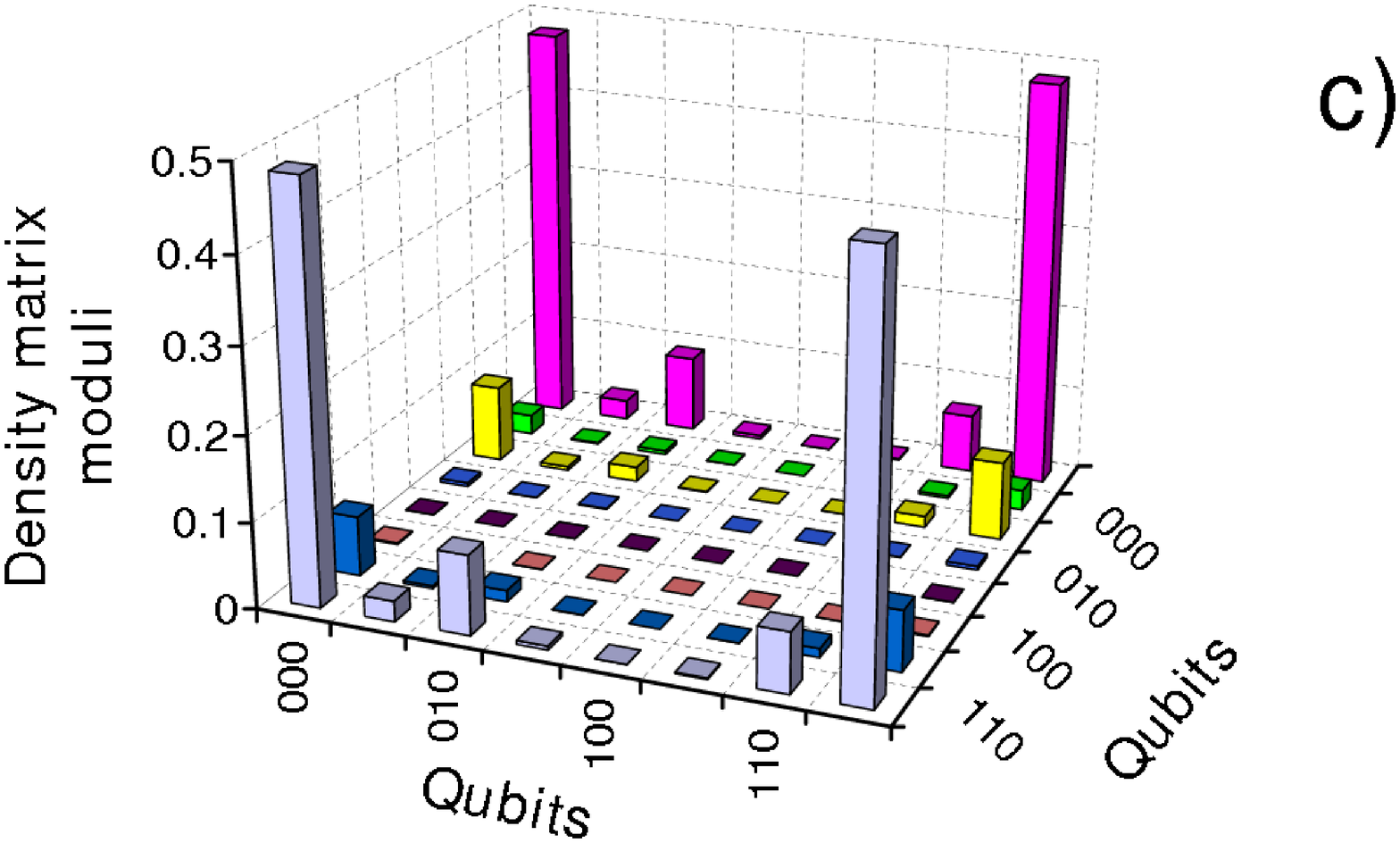}
     \end{minipage}
     \caption{\label{figden11} (Color online). The density matrix of the qubits
       $\{x_0,y_1,y_3\}$ evaluated at three different time steps: a)
       input, b) initialization procedure, and c) after the modular
       exponentiation. Note that these matrices have been obtained
       from the full density matrices of the electrons of the qubits
       $\{x_0,y_1,y_3\}$ by integrating over the variables defining
       the position of three carriers along the channels. Here the
       moduli of the density matrix elements are plotted.}
    
   \end{center}
 \end{figure}

 Figure~\ref{figden11} shows the density matrix describing the qubits
 of the argument and function register,$\{x_0,y_1,y_3\}$ at three
 different stages: input, initialization procedure and modular
 exponentiation. For sake of simplicity, we do not consider the $x_1$,
 $y_0$, and $y_2 $ qubits that evolve trivially during the computation.
 The knowledge of the joint state of both registers after modular
 exponentiation is essential for the estimation of the device
 performance. In particular, we find that the output quantum state
 corresponds to the GHZ-like entangled state $|\Psi\rangle$ of
 Eq.~(\ref{GHZ2}) with a good approximation. For a more quantitative
 evaluation of the reliability of the algorithm, we have also
 calculated the fidelity $F=\langle \Psi|\rho_{out}|\Psi \rangle$,
 where $\rho_{out}$ is the output density matrix of the full system.
 The high value found, $F$=0.97, evidences the very good efficiency of
 the implementation. Such a result is certainly related to the fact
 that our numerical simulations have been performed by setting the
 device temperature at 0 K, that is neglecting any effect of
 decoherence induced by the environment on the carrier transport
 properties. In particular, this means that the electron-phonon
 interaction have not been included in the simulations. These
 assumptions are physically sound when we take into account that the
 experimental investigations of the low-dimensional structures used as
 the basic blocks for our device are usually performed at very low
 temperatures~\cite{Bird, Barnes2}. Moreover, one of the pros of our
 results certainly relies on the high fidelity value, which has not
 been obtained under ideal geometries for the $R_0(1)(\phi)$ and
 $T(\gamma)$ gates. Once more, this off-ideality situation is close to
 the experimental conditions.


\begin{figure}[htpb]
   \begin{center}
     \includegraphics*[width=0.5\linewidth]{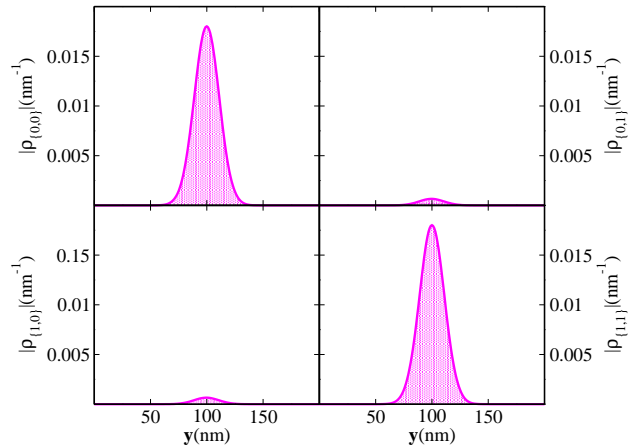}
     \caption{\label{figwav} (Color online). Density matrix
       $\rho_{\{x_0,x_0^{\prime}\}}(\mathbf{y}, \mathbf{y})$ of the
       electron of qubit $x_0$ after the modular exponentiation. The
       diagonal elements describe the density probability of finding
       the electron in the point $\mathbf{y}$ along the wire 0 or 1.
       Here the moduli of $\rho_{\{x_0,x_0^{\prime}\}}(\mathbf{y},
       \mathbf{y})$ are plotted. Note that the curves reported in the left panels refer to the left ordinate axes, while the ones reports in the right panels refer to the right ordinate axes.}
   \end{center}
 \end{figure}

 The density matrix of the argument register after modular
 exponentiation is displayed in Fig.~\ref{figwav}. Specifically, we
 show the reduced density matrix
 $\rho_{\{x_0,x_0^{\prime}\}}(\mathbf{y}, \mathbf{y})$ of the electron
 of qubit $x_0$, without the redundant qubit $x_1$. The output of the
 quantum circuit is the logical state probability, that is the
 probability of finding the electron of the qubit $x_0$ in the wire 0
 or 1. This is described by the integral over $\mathbf{y}$ of the
 diagonal elements of
 $\rho_{x_0,x_0^{\prime}}(\mathbf{y},\mathbf{y})$. The off-diagonal
 terms are very small, thus proving that the argument register becomes
 a full quantum statistical mixture because of its entanglement with
 the function register. To better quantify the amount of the quantum
 correlations created between $x_0$ and $\{y_1,y_3\}$, we evaluate the
 linear entropy $\varepsilon_{L}$ of the qubit $x_0$
 as~\cite{Buscemi3}:
 $\varepsilon_{L}=2\left(1-\mathrm{Tr}\rho_r^2\right)$, where the
 factor 2 stems out from the normalization condition and $\rho_r^2$ is
 the square of the reduced density matrix
 $\rho_{\{x_0,x_0^{\prime}\}}(\mathbf{y},\mathbf{y})$ of the electron
 of the qubit $x_0$ integrated over $\mathbf{y}$. We find that
 $\varepsilon_{L}=0.999$ and therefore a maximal correlation between
 the two registers of the quantum circuit is build up, that
 unambiguously proves the quantum nature of the simulated circuit, as
 required by Shor's algorithm.  Once the logical state probabilities
 of the qubit $x_0$ are known, the latter are combined with the qubit
 $x_1$ in the zero state and then, as required, the order of the
 argument bits is inverted. This procedure allows one to obtain the
 binary output of the circuit already discussed in Sect.~\ref{c11},
 namely 00 and 10. The first is found with a probability of 50,1\% and
 represents the expected failure of the Shor's algorithm, whereas the
 second is obtained with a 49,9\% probability and leads a to
 successful determination of the order $r$,. As theoretically
 expected~\cite{Shor1,Beckman}, failure and success have equal
 probabilities. These outputs indicate an almost ideal performance of
 the quantum algorithm.

\subsection{C=2}
\begin{figure}[b]
  \centering
  \begin{minipage}[htbp]{\textwidth}
    \includegraphics*[width=0.3\linewidth]{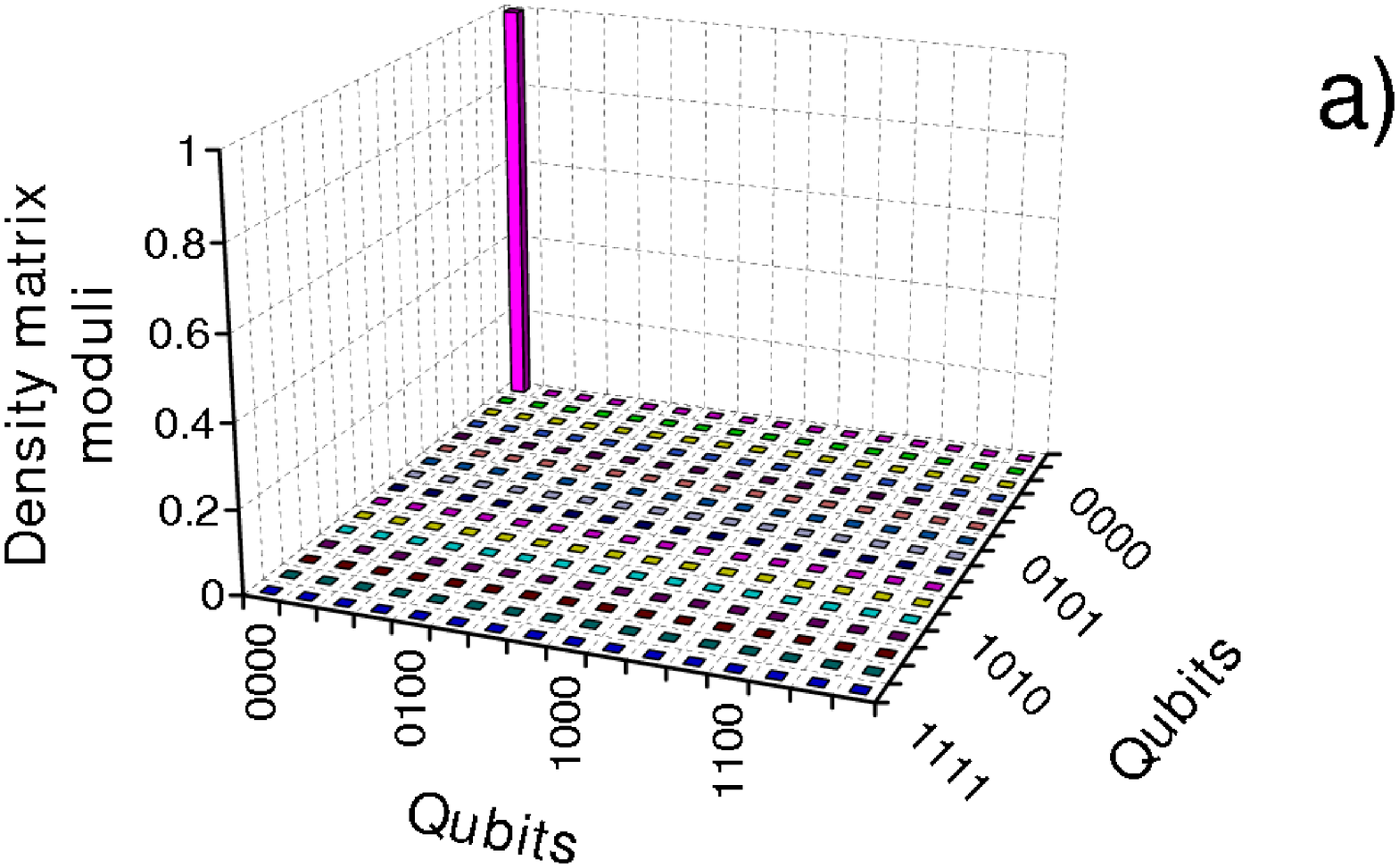}
    \includegraphics*[width= 0.3\linewidth]{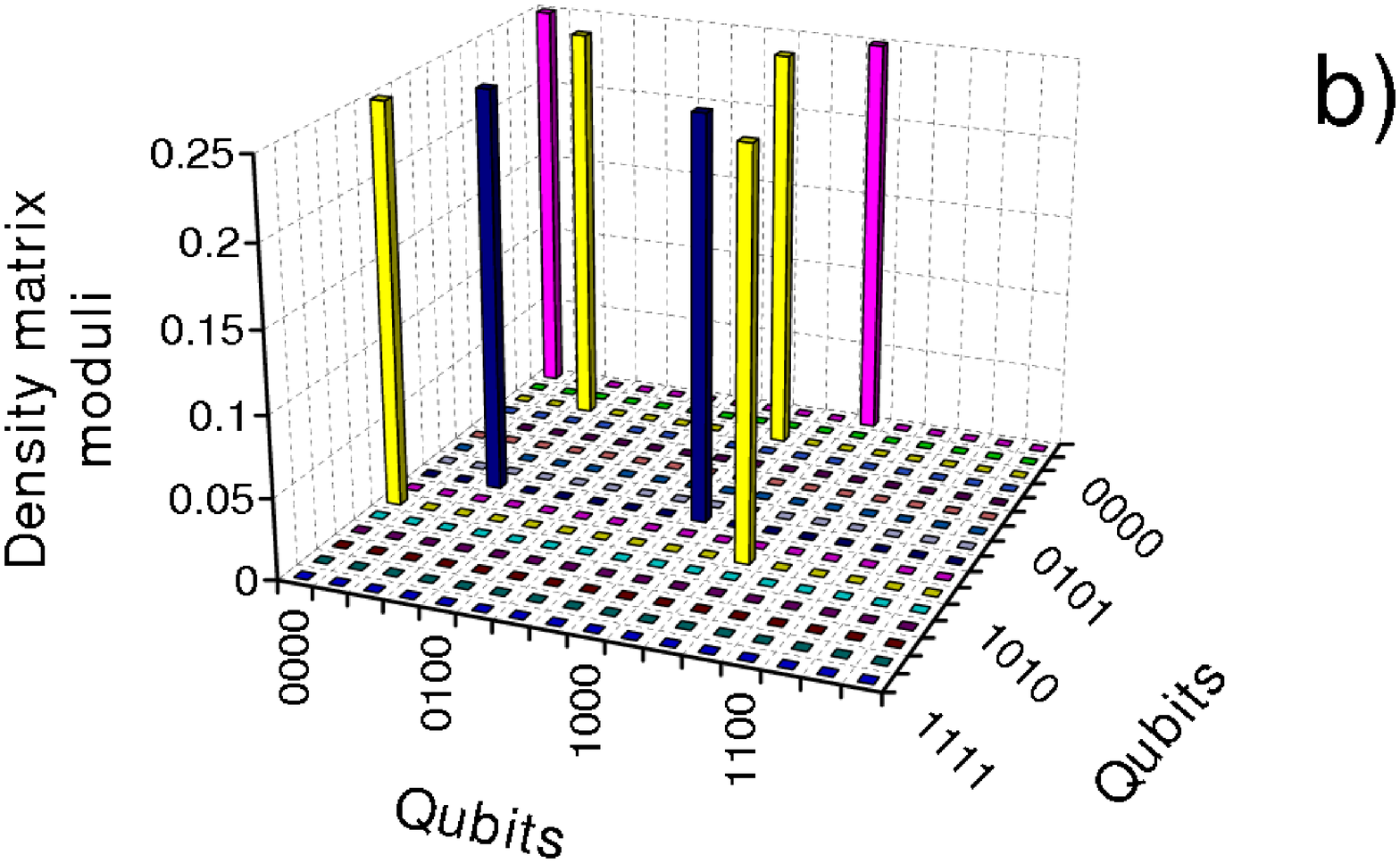}
    \includegraphics*[width= 0.3\linewidth]{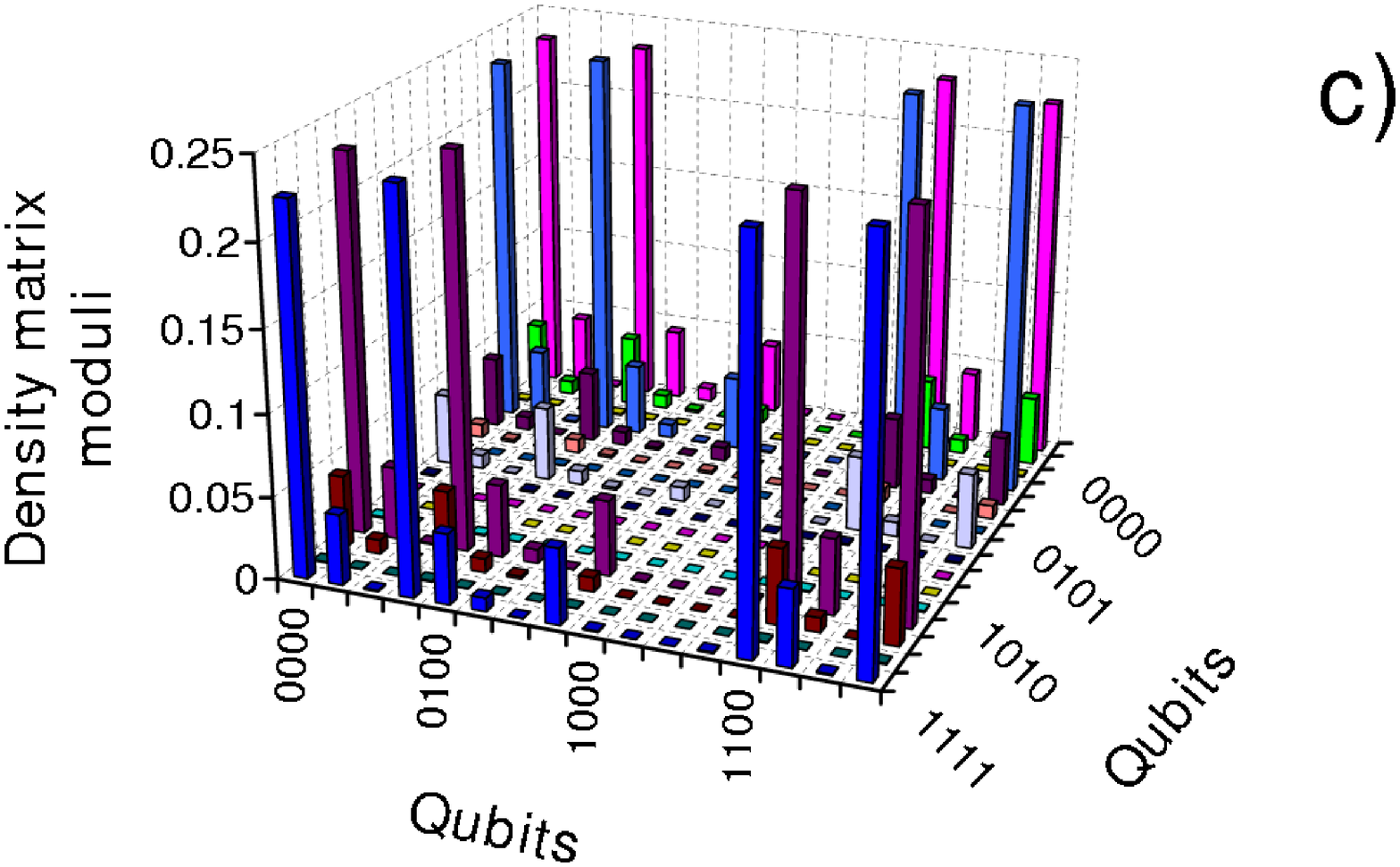}
  \end{minipage}
  \caption{\label{figden2} (Color online). The density matrix of the qubits
    $\{x_0,x_1,y_0,y_1\}$ evaluated at three different time steps: a)
    input, b) initialization procedure, and c) after the modular
    exponentiation. Note that these matrices have been obtained from
    the full density matrices of the electrons of the qubits
    $\{x_0,x_1,y_0,y_1\}$ by integrating over the variables defining
    the position of three carriers along the channels. Here the moduli
    of the density matrix elements are plotted.}
\end{figure}

\begin{figure}[htpb]
  \begin{center}
    \includegraphics*[angle=270,width=\linewidth]{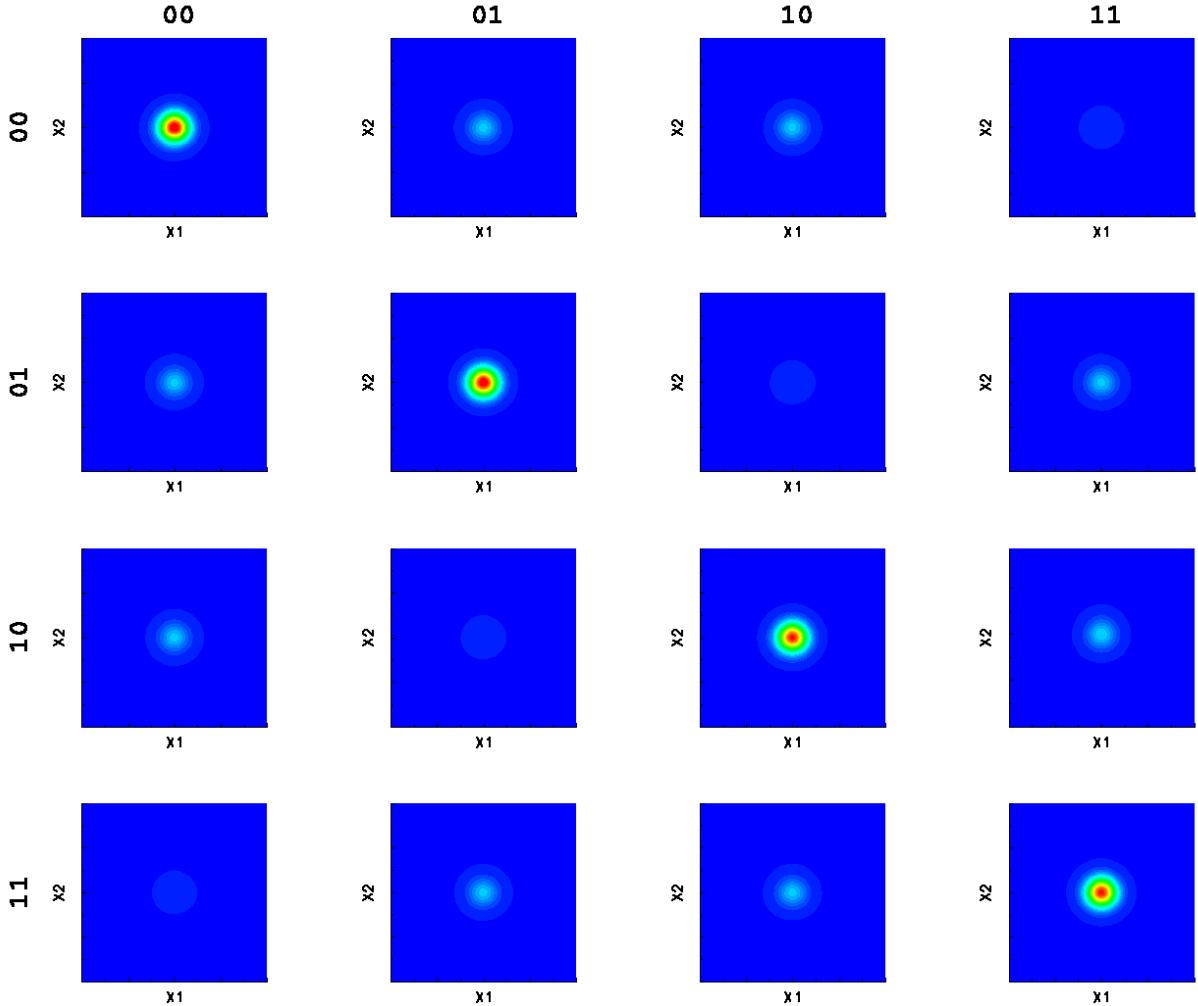}
    \caption{\label{figwav2} (Color online). Density matrix
      $\rho_{\{x_0,x_0^{\prime},x_1,x_1^{\prime}\}}(\mathbf{y},
      \mathbf{y},\mathbf{y}^{\prime},\mathbf{y}^{\prime})$ of the
      couple of electrons of qubit $\{x_0,x_1\}$ after the modular
      exponentiation. The diagonal elements describe the density
      probability of finding the two electrons in the points
      $\mathbf{y}$ and $\mathbf{y}^{\prime}$ along the wire 0 or 1.
      Here the moduli of
      $\rho_{\{x_0,x_0^{\prime},x_1,x_1^{\prime}\}}(\mathbf{y},
      \mathbf{y},\mathbf{y}^{\prime},\mathbf{y}^{\prime})$ are
      plotted.}
  \end{center}
\end{figure}

The numerical investigation of the compiled version of Shor's
algorithm with $C$=2 and the evaluation of the function
$\ln_2[2^{x}\!\!\!\!\!\mod 15]$ in the function register required the
evaluation of the time evolution of all of the qubits of the
registers: $x_0$, $x_1$, $y_0$, $y_1$. The density matrix describing
the global system is displayed in Figure~\ref{figden2} at three
different time steps. The output quantum state describes, with a
fidelity of 0.89, the product of two maximally entangled Bell pairs,
as theoretically expected.
The argument register outputs are reported in Figure~\ref{figwav2},
where the reduced density matrix
$\rho_{\{x_0,x_0^{\prime},x_1,x_1^{\prime}\}}(\mathbf{y},
\mathbf{y},\mathbf{y}^{\prime},\mathbf{y}^{\prime})$ of the couple of
the electrons of qubits $\{x_0,x_1\}$ is displayed. The argument
register is almost maximally mixed as a consequence of the
entanglement with the qubits $\{y_0,y_1\}$, as the large value of the
linear entropy $\varepsilon_{L}=0.976$ confirms. The binary output of
the algorithm, namely one among the possible two-bit responses 00, 01,
10 and 11, is obtained by considering the probabilities of the logical
state of the qubits $\{x_0,x_1\}$ and then inverting their order. The
second and the fourth terms yield $r$=4, which gives correctly the
factors 3 and 5 once processed in the classical Euclid's algorithm; on
the contrary, the first value corresponds to a failure mode whereas
the third on leads to trivial factors. All of the outcomes have
exactly the same probability to happen, which consequently means that
the routine has a success rate of 50\% like the previously-discussed
case of $C$=11.

Although the binary outputs of the argument registers clearly indicate
that the algorithm performance is near-ideal both for $C$=11 and
$C$=2, the efficiency of the quantum networks is slightly different in
the two implementations investigated, as evidenced by the estimated
fidelity and degree of entanglement between the registers. More
specifically, the quantum circuit performance gets worse moving from
$C$=11 to $C$=2. Such a behavior, at a first sight, appears to be
surprising, due to the larger number of one- and two qubits quantum
gates numerically simulated in the former implementation.  A possible
explanation could amazingly bring up the small errors inherent the
tuning of the quantum gates. In fact, the flaws of the one -and
two-logical operations could counterbalance each other resulting as a
net effect a higher efficiency of quantum circuits.

\section{Conclusions}\label{Conclu}
Shor's algorithm highlights the potential power of quantum computation
and nowadays its realization in scalable structures represents one of
the main challenges of quantum information science. Only in recent
years experimental demonstrations of this algorithm have been given in
some physical scenarios ranging from NMR to photon qubits.
Nevertheless, the quantum nature of the processes and/or the
scalability of the investigated systems in these approaches are
questionable.

In this paper we have introduced and numerically simulated an
implementation of the easiest meaningful example of the Shor's
algorithm, that is the factorization of $N$=15, through co-primes
$C$=11 and $C$=2.  The idea we have proposed exploits the coherent
SAW-assisted transport of electrons in networks of coupled quantum
wires, and has a great potential in view of its integrability with
conventional microelectronics and of its scalability to more complex
systems containing many qubits. Quantum information is processed by
means of a sequence of one- and two-qubit gates,
 materialized by means of an electronic beam splitter and 
phase shifter and a Coulomb coupler,
respectively.  Their  experimental realization 
in semiconductor quantum wires is very challenging
since it requires the use of frontier  semiconductor
technology. Only in the last years, prototype blocks
mimicking single-qubit rotations in a couple of  1D channels
have been experimentally demonstrated~\cite{Bird,Fischer3,Fischer2}. 
In particular,
the  switching of coupled-quantum wire  qubit
characteristics has been  explored~\cite{Bird}. Furthermore, 
Fischer \emph{et al.} controlled the coupling between two modes of 
a couple of 1D channels, obtained exploiting the two minima of 
the conduction-band edge in the growth direction
of a GaAs 2D electron gas~\cite{Fischer3}  or two vertically-coupled 
2D electron gases~\cite{Fischer2}. No experimental evidence of 
two-qubit operation in quantum wires networks has been achieved so far.
On the other hand, the coherent manipulation
of charge states in two spatially separated double quantum dots integrated
in a GaAs/AlGaAs heterostructure has been realized~\cite{Haya, Shink}. 
Specifically, two-qubit operations (swap and controlled-rotation) 
have been successfully implemented.
  
We stress that the protocol here proposed for the order-finding
routine at the heart of Shor's algorithm represents a``'non standard''
implementation of the quantum circuits as commonly used in the
literature for quantum factorization~\cite{Shor1,Ekert}. Such an
implementation keeps the basic features of the original algorithm
(i.e., ``massive parallelism'' given by the entanglement between the
quantum registers and binary output), and also allows for a simple
network with a lower number of fundamental gates.  This makes the
numerical simulation of the presented protocol less demanding and
could also have interesting perspectives on the full-scale realization
of Shor's algorithm.

The high efficiency of the quantum processes simulated is shown by the
large values obtained for fidelities. Furthermore, also the success
rate of the algorithm are close to its ideal value, in agreement with
recent experimental investigations~\cite{Lanyon}. The algorithm
performance is even more noteworthy if we do consider the good but not
ideal geometry of the logic gates and compare our data with those of
the near-ideal case. This behavior is a clear signature of the
robustness of the algorithm, which is also able to accommodate small,
but non negligible errors coming from the fabrication and tuning of
the quantum gates. The capability to take into account small
deviations from ideality is certainly a plus, that makes the algorithm
to compare favourably to any of its experimental implementations. In
fact, it gives the opportunity to let the device work correctly even
in presence of unavoidable environmental decoherence effects, always
present even at low temperatures.

Since the recent developments in nanostructure fabrication opened new
scenarios in scalable electronic quantum
computation~\cite{Barnes2,Fischer1,Fischer2}, the promising results
here presented indicate a fruitful guideline for the research in
quantum information science. Specifically, this work highlights a
peculiar physical architecture which could become, in a near future, a
powerful mean to implement a broader variety of quantum algorithms and
therefore to fully exploit the whole potential of quantum computation.
 
\begin{acknowledgments}
  The author would like to thank E.~Piccinini, P.~Bordone, A.~Bertoni,
  and C.~Jacoboni for stimulating discussions and valuable help.
\end{acknowledgments}


\begin{thebibliography}{33}
\bibitem[{NOT()}]{NOTE3}
\bibinfo{note}{In order to make less challenging the computational procedure,
  in our analysis we have not taken into account the qubits $y_0$ and $y_2$
  evolving trivially.}
\end{thebibliography}
\end{document}